\documentclass[a4paper,conference]{IEEEtran}

\usepackage{cite}

\usepackage{tikz}
\usetikzlibrary{arrows,snakes,shapes,positioning,arrows,decorations.markings,calc}
\usetikzlibrary{dsp,chains}
\usepackage{psfrag}
\usepackage{color}
\definecolor{green}{rgb}{0,0.4,0.05}
\definecolor{red}{rgb}{0.8,0,0}

\usepackage{amsmath}

\usepackage{algorithm}
\usepackage{algpseudocode}
\newcommand{\vect}[1]{\mathbf{#1}}
\newcommand{\matt}[1]{\mathbf{#1}}
\newcommand{\jim}{\mathrm{j}\,}

\DeclareMathOperator*{\E}{E}
\newcommand{\T}{\operatorname{\mathrm{T}}}
\newcommand{\Q}{\operatorname{\mathcal{Q}}}
\newcommand{\He}{\text{H}}
\newcommand{\diag}{\text{diag}}
\newcommand{\adj}{\text{adj}}
\newcommand{\sign}{\text{sign}}

\begin{document}

\title{Minimum BER Precoding \\in
1-Bit Massive MIMO Systems}

\author{\IEEEauthorblockN{Hela Jedda and Josef~A.~Nossek}
\IEEEauthorblockA{Institute of Circuit Theory and Signal Processing\\
Technische Universit\"at M\"unchen, 80290 Munich Germany\\
Email: \{hela.jedda, josef.a.nossek\}@tum.de}
\and
\IEEEauthorblockN{Amine Mezghani}
\IEEEauthorblockA{University of California, Irvine\\
Irvine, CA 92697, USA\\
Email: amezghan@uci.edu}}

\maketitle
\tikzset{DSP lines/.style={help lines,very thick,color=black}}
\tikzset{line_arrow/.style={help lines,very thick,color=black,->,-angle 90}}
\tikzset{filter/.style={rectangle,inner sep=0pt,minimum height=0.8cm,minimum width=1cm,draw=black,very thick}}
\tikzset{delay/.style={rectangle,inner sep=0pt,minimum size=1cm,draw=black,very thick}}
\tikzset{downsampling/.style={rectangle,inner sep=0pt,minimum height=0.8cm,minimum width=0.7cm,draw=black,very thick}}
\tikzset{upsampling/.style={rectangle,inner sep=0pt,minimum height=0.8cm,minimum width=0.7cm,draw=black,very thick}}
\tikzset{empty_node/.style={inner sep=0pt,minimum size=0cm}}
\tikzset{connection/.style={circle,draw=black,fill=black,inner sep=0pt,minimum size=2mm}}
\tikzset{coefficient/.style={isosceles triangle,draw=black,very thick,inner sep=0pt,minimum size=.7cm}}
\tikzset{source/.style={semicircle,minimum size=.5cm,draw=black,very thick,shape border rotate=270}}
\tikzset{adder/.style={circle,minimum size=.25cm,inner sep=0pt,draw=black,very thick}}
\tikzset{multiplier/.style={circle,minimum size=.25cm,inner sep=0pt,draw=black,very thick}}
\tikzset{double_arrow/.style={double distance=5pt,thick,shorten >= 6pt,decoration={markings,mark=at position 1 with {\arrow[scale=.6,>=angle 90]{>}}},postaction={decorate}}}
\begin{abstract}
1-bit digital-to-analog (DACs) and analog-to-digital converters (ADCs) are gaining more interest in massive MIMO systems for economical and computational efficiency. We present a new precoding technique to mitigate the inter-user-interference (IUI) and the channel distortions in a 1-bit downlink MU-MISO system with QPSK symbols. The transmit signal vector is optimized taking into account the 1-bit quantization. We develop a sort of mapping based on a look-up table (LUT) between the input signal and the transmit signal. The LUT is updated for each channel realization. Simulation results show a significant gain in terms of the uncoded bit-error-ratio (BER) compared to the existing linear precoding techniques.
\end{abstract}

\IEEEpeerreviewmaketitle

\section{Introduction}
Massive MIMO systems have been seen as a promising technology for the next generation wireless communication systems \cite{Marzetta2010, Bjornson2013}. The huge increase in the number of antennas at the base station (BS) can improve spectral efficiency (SE), energy efficiency (EE) and reliability.
The BS with large number of antennas, say 100 antennas or more, serves simultaneously 
a much smaller number of single-antenna users. 
The price to pay for massive MIMO systems is increased complexity of the hardware (number of radio frequency (RF) and ADC/DAC chains) and the signal processing and resulting increased energy consumption at the transmitter \cite{Rusek2013}. Several approaches are considered in the literature 
to decrease the power consumption such as spatial modulation \cite{Renzo2014,MullerSedaghat}, the use of parasitic antennas \cite{Kalis2012,Sedaghat} and the use of low-cost transceivers \cite{Bjornson2015}. 
One attractive solution  to overcome the issues of high complexity and high energy consumption associated with massive MIMO, 
is the use of very low resolution ADCs and DACs. The power consumption of the ADC and the DAC, one of the most power-hungry devices, can be reduced exponentially by decreasing the resolution \cite{Svensson2006} and 1-bit quantization can
drastically simplify other RF-components, e.g., amplifiers and mixers. 
Therefore, we focus on massive MIMO systems where the resolution of the DACs and ADCs is restricted to 1 bit, e.g. 1-bit massive MIMO systems.

With the knowledge of CSI at the BS (CSIT), this large spatial DoF of massive MIMO systems can be exploited to significantly increase the spatial multiplexing/diversity gain using MU-MIMO precoding \cite{Peel2005, Gershman2010}. In the literature \cite{Mezghani2009, Usman2016} linear precoders are designed for 1-bit massive MIMO systems based on minimum-mean-square-error criterion (MMSE) to mitigate IUI and the distortions due to the coarse quantization. However, the MMSE criterion may be not optimal since the desired receive signals are restricted to discrete QPSK points and they can be corectly detected if they belong to the respective quadrants. Thus, we aim at changing the design criterion to the minimun BER (MBER). The goal is to get the receive signal in the desired quandrant and as far as possible from the decision thresholds.

In this contribution, we do not design a precoder but the transmit signal vector. So, we design a sort of mapping based on a LUT between the input signal vector and the transmit signal vector. The LUT is updated for each channel. The entries of the transmit signal vector belong to the square formed by the QPSK constellation points to minimize the quantization distortions at the transmitter.

This paper is organized as follows: in Section \ref{sec:sysmodel} we introduce the downlink MU-MISO system model. In Section \ref{sec:mapping} we give an overview about the mapping idea. The MBER criterion is illustrated and explained in Section \ref{sec:mber}. In Section \ref{sec:optproblem} we formulate the optimization problem based on the MBER criterion and show the derivations and the corresponding solution. We give two linear precoder designs in Section \ref{sec:linear_precoders}, that we aim at comparing with. In Sections \ref{sec:simresults} and \ref{sec:conc} we interpret the simulation results and summarize this work.

Notation: Bold letters indicate vectors and matrices, non-bold letters express scalars. The operators $(.)^{*}$, $(.)^{\rm T}$, $(.)^{\rm H}$, ${\rm adj}(.)$  and $\E\left[.\right]$ stand for complex conjugation, the transposition, Hermitian transposition, adjugate and the expectation, respectively. The $n \times n$ identity matrix is denoted by $\mathbf{I}_{n}$ while the zeros (ones) matrix with $n$ rows and $m$ columns is defined as $\mathbf{0}_{n,m}$ ($\mathbf{1}_{n,m}$). The vector $\vect{e}_{l}$ represents a zero vector with 1 in the $l$-th position. We define $\left(\bullet\right)_{R} = \Re \lbrace \bullet \rbrace$, $\left(\bullet\right)_{I} = \Im \lbrace \bullet \rbrace$ and $\mathcal{Q}(x) = 1/ \sqrt{2} \left( \sign(x_{R}) + \jim \sign(x_{I})\right) $ with $\sign(0) =1$. Additionally, $\diag(\matt{A})$ denotes a diagonal matrix containing only the diagonal
elements of $\matt{A}$. 
\section{System Model}
\label{sec:sysmodel}
\begin{figure}[h]
\centering
\resizebox{9cm}{!} {%
\begin{tikzpicture}

\node (in){};
\node[filter] (mapping) [right=of in] {$\mathcal{M}$};
%\node[filter] (lut) [below=of mapping][yshift=0.5cm] {LUT};
\node[filter] (quantizer1) [right=of mapping] {$\mathcal{Q} (\bullet) $};
\node[filter] (channel) [right=of quantizer1]  {$\matt H$};
\node[adder] (noise) [right=of channel]  {$+$};
\node (n)[below=of noise][yshift=0.5cm] {$\boldsymbol{\eta}$};
\node[filter] (quantizer2) [right=of noise][xshift=-0.5cm] {$\mathcal{Q} (\bullet) $};
\node (out) [right=of quantizer2]{};

\draw[DSP lines] [-stealth] (in.east) -- (mapping.west)node[pos=0.5,above]{$\vect s$}node[pos=0.5,below]{$\mathcal{O}^{M\times 1}$} ;
\draw[DSP lines] [-stealth] (mapping.east) -- (quantizer1.west)  node[pos=0.5,above]{$\vect x$} node[pos=0.5,below]{$\mathcal{C}^{N\times 1}$} ;
%\draw[DSP lines] [dashed] (lut.north) -- (mapping.south);
\draw[DSP lines] [-stealth] (quantizer1.east) -- (channel.west)node[pos=0.5,above]{$\vect x_Q$}node[pos=0.5,below]{$\mathcal{O}^{N\times 1}$};
\draw[DSP lines] [-stealth] (channel.east) -- (noise.west) node[pos=0.5,above]{$\vect { r}$}node[pos=0.5,below]{$\mathcal{C}^{M\times 1}$};
\draw[DSP lines] [-stealth] (n) -- (noise.south);
\draw[DSP lines] [-stealth] (noise.east) -- (quantizer2.west);
\draw[DSP lines] [-stealth] (quantizer2.east) -- (out) node[pos=0.5,above]{$\vect {\hat s}$}node[pos=0.5,below]{$\mathcal{O}^{M\times 1}$}  ;

\end{tikzpicture}%
}
\caption{Downlink MU-MISO System model with QPSK symbols}
\label{fig:sys_model}
\end{figure}
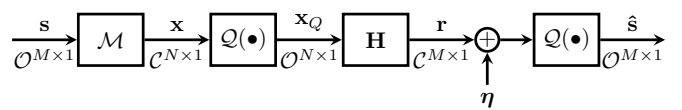

We consider a 1-bit downlink MU-MISO scenario as depicted in Fig. \ref{fig:sys_model}. The BS with $N$ antennas serves $M$ single-antenna users, where $N \gg M$. The signal vector $\vect{s} \in \mathcal{O}^{M}$ contains the data symbols for each of the $M$ users, where $\mathcal{O}$ represents the set of QPSK constellation. We assume that $\vect{s}\sim \mathcal{O}\left ( \vect{0}_M,\sigma_{s}^2\matt{I}_{M} \right )$. In this system we deploy 1-bit quantization $\mathcal{Q}$ at the transmitter as well as at the receiver. The use of the 1-bit quantizer at the transmitter delivers a signal $\vect{x}_{Q} \in \mathcal{O}^N$. To mitigate the IUI and the distortions due to the coarse quantization, the input signal vector $\vect s$ is mapped to the unquantized transmit signal vector $\vect x$ prior to DAC. This mapping is based on a LUT of size $N\times 4 ^M$, that is generated at the beginning of each coherence slot. 
The transmit signal gets scaled with $\sqrt{\frac{E_{\text{tx}}}{ N}}$, where $E_{\text{tx}}$ is the available power at the transmitter.
The received decoded signal vector $\hat{\vect{s}} \in \mathcal C^{M \times 1}$ of the $M$ single-antenna users reads as $\hat{\vect{s}} = \Q \left(\sqrt{\frac{E_{\text{tx}}}{N}} \matt H \vect{x}_{Q} + \boldsymbol{\eta}\right)$, where $\matt H \in \mathcal{C}^{M\times N}$ is the channel matrix with i.i.d. complex-valued entries of zero mean and unit variance and $\vect{\boldsymbol{\eta}}\sim \mathcal{C}\mathcal{N}\left ( \vect{0}_M,\matt{C}_{\boldsymbol{\eta}} = \matt{I}_{M} \right )$ is the AWG noise vector.

\section{Mapping: $\mathcal{M}$}
\label{sec:mapping}
\begin{figure}[h]
\centering  
\psfrag{Nx4M}[][]{$N\times 4^M$}
\includegraphics[width=0.4\textwidth]{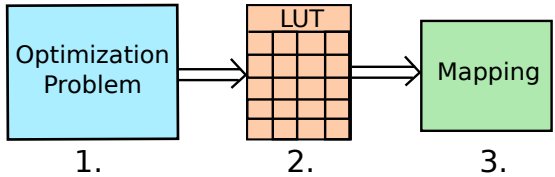}
\caption{Processing steps for each channel}
\label{fig:processing_steps}
\end{figure}

In this work, we do not design a precoder but we design the transmit vector signal $\vect x$ for a given input signal vector $\vect s$ depending on the channel, while we assume full CSIT. As depicted in Fig. \ref{fig:processing_steps}, first, an optimization problem is solved for all possible input vectors $\vect s$ to find the optimal transmit vectors $\vect x$. The used optimization problem is introduced in Section \ref{sec:optproblem}. Second, the solutions are stored in the LUT of size $N \times 4^M$. Since we are restricted to QPSK modulation, we get $4^M$ possible input vectors. Third, we map the given input vector $\vect s$ into a signal vector $\vect x$ according to the LUT, which is updated for each channel.

The aim of the optimization problem is to jointly minimize the IUI and the quantization distortions. The optimization criterion is the minimun BER (MBER) under the constraint that $\vect x \in \mathcal{O}^N$. This constraint leads to a linear behavior of the quantizer at the transmitter, e.g. $\vect x_Q = \vect x$. Thus, the quantization distortions at the transmitter are omitted.

\section{MBER criterion}
\label{sec:mber}
\subsection{Single User Scenario}
\begin{figure}[h]
\centering  
\psfrag{Re}[][]{$\Re$}
\psfrag{Im}[][]{$\Im$}
\includegraphics[width=0.25\textwidth]{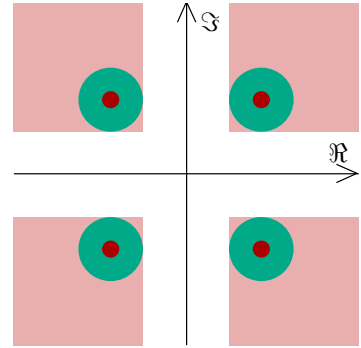}
\caption{MMSE vs. MBER criterion}
\label{fig:mmse_vs_ber}
\end{figure}

To minimize the BER in the case of QPSK symbols, we need to get the receive signal in the same quadrant as the desired signal. Since the receive signal gets distorted with some additive Gaussian noise that may remove it from the desired quadrant, we need to get the receive signal as far as possible from the quantization thresholds to make it less sensitive to the noise. For illustration we consider Fig. \ref{fig:mmse_vs_ber}. The red points designate the QPSK constellation. The solution set for the MBER criterion is represented by the four half-bounded squares. However, the MMSE criterion tries to have the receive signal as close as possible to the desired signal. So we get the green circles. Thus, the MMSE solution set is restricted to a subset of the MBER solution set. When massive MIMO is employed, the signals can get larger magnitude and this is prohibited by MMSE but preserved by MBER. 

\begin{figure}[h]
\centering  
\psfrag{Re}[][]{$\Re$}
\psfrag{Im}[][]{$\Im$}
\psfrag{s}[][]{$\!\!s$}
\psfrag{shat}[][]{$r$}
\psfrag{shatr}[][]{$r_{\Re}$}
\psfrag{shati}[][]{$r_{\Im}$}
\psfrag{phi}[][]{$\phi$}
\psfrag{A}[][]{\textcolor{green}{$\Re\{r s^*\}$}}
\psfrag{B}[][]{\textcolor{red}{$\Im\{r s^*\}$}}
\includegraphics[width=0.25\textwidth]{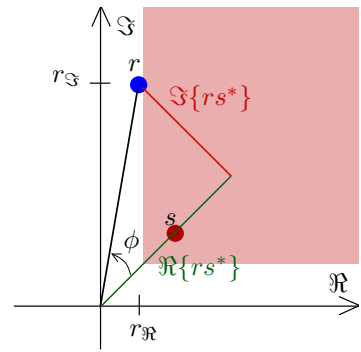}
\caption{Illustration of the optimization problem}
\label{fig:problem_illustration}
\end{figure}

In order to explore the MBER criterion, we have to formulate an appropriate mathematical optimization problem. To this end we refer to Fig. \ref{fig:problem_illustration} for illustration. As mentioned above we aim at making the receive signal $r$ belong to the safe red area. One way is to maximize $\Re\{r s^*\}$ and minimize $\vert \phi \vert$, where $\phi \in \left]-\frac{\pi}{4}, \frac{\pi}{4} \right[ $. Fortunately, there is a mathematical expression that can enable maximizing $\Re \{r s^*\}$ while $\phi \in \left]-\frac{\pi}{4}, \frac{\pi}{4} \right[ $, which is given by
\begin{align}
\max_{\vect x} \Re \{\left( r s^* \right)^2\} &=  \max_{\vect x} \vert r \vert^2 \vert s \vert ^2 \cos(2 \phi) \nonumber \\
&\text{s.t. } \vect x \in \mathcal{O}^N.
\label{eq:opt_prob_su}
\end{align}
The solution of (\ref{eq:opt_prob_su}) requires that $\cos(2\phi)$ is positive which is achieved by $\phi \in \left]-\frac{\pi}{4}, \frac{\pi}{4} \right[$.
\subsection{Multi User Scenario}
For the multi user scenario we make use of the same optimization problem in (\ref{eq:opt_prob_su}) and apply it for each user
\begin{align}
\max_{\vect x} \Re \{\left(  r_m s_m^* \right)^2\} &= \max_{\vect x} \vert r_m \vert^2 \vert  s_m \vert^2 \cos(2 \phi_m), m=1,2,...,M \nonumber \\
&\text{s.t. } \vect x \in \mathcal{O}^N,
\label{eq:opt_prob}
\end{align}
where $\vect r = \sum_{m=1}^{M} r_m \vect e_m$ and  $\vect s = \sum_{m=1}^{M} s_m \vect e_m$.
The $M$ cost functions can be jointly expressed by the following matrix 
\begin{align}
\matt P &= \Re\left\lbrace \diag\left(\vect{r} \vect{s}^{\He} \right)^2 \right\rbrace \nonumber \\ 
&= \Re\left\lbrace \diag\left(\matt H \vect{x} \vect{s}^{\He} \right)^2 \right\rbrace \nonumber \\
&=\! \diag \left( \underbrace{\begin{bmatrix} \matt {H_{\Re}} \!&\! -\matt {H_{\Im}}
\end{bmatrix}}_{\matt C} \underbrace{\begin{bmatrix}
\vect {x_{\Re}} \\ \vect {x_{\Im}}\end{bmatrix}}_{\vect x'} \vect{s_{\Re}}^{\T} \!+\!\underbrace{\begin{bmatrix} \matt {H_{\Im}} \!&\! \matt {H_{\Re}}
\end{bmatrix}}_{\matt D} \begin{bmatrix}
\vect {x_{\Re}} \\ \vect {x_{\Im}}\end{bmatrix} \vect{s_{\Im}}^{\T}\right)^2 \nonumber \\
& -\! \diag \left( \begin{bmatrix} \matt {H_{\Re}} \!&\! -\matt {H_{\Im}}
\end{bmatrix} \begin{bmatrix}
\vect {x_{\Re}} \\ \vect {x_{\Im}}\end{bmatrix} \vect{s_{\Im}}^{\T}\!-\!\begin{bmatrix} \matt {H_{\Im}} \!&\! \matt {H_{\Re}}
\end{bmatrix} \begin{bmatrix}
\vect {x_{\Re}} \\ \vect {x_{\Im}}\end{bmatrix} \vect{s_{\Re}}^{\T}\right)^2 \nonumber \\
&= {\underbrace{\diag(\matt C \vect x' \vect s_{\Re}^{\T} + \matt D \vect x' \vect s_{\Im}^{\T})}_{\matt A}}^2 -{\underbrace{\diag(\matt C \vect x' \vect s_{\Im}^{\T} - \matt D \vect x' \vect s_{\Re}^{\T})}_{\matt B}}^2 \nonumber \\
&= \matt A^2 - \matt B^2.
\end{align} 

We end up with $M$ cost functions that have to be jointly maximized with a single transmit vector $\vect x$. These $M$ cost functions have to be combined to maximize them together. Here the question arises: how to do that? Shall we maximize the sum, the minimal contribution or the product?

\section{Optimization Problem}
\label{sec:optproblem}
The optimization problem can not be convex since the solution set of $\vect x = \sum_{n=1}^{N} x_n \vect e_n \in \mathcal{O}^N$ is non-convex. The optimal solution can be found by exhaustive search. However, the complexity of the exhaustive search increases exponentially with the number of antennas $N$. To decrease the complexity of the problem and make it solvable with linear methods we relax the constraint to $\vert{\Re\{ x_n\}}\vert \leq 1/\sqrt{2}$ and $\vert{\Im\{ x_n\}}\vert \leq 1/\sqrt{2},$ $n =1,2,...N$. Since the matrix $\matt P$ is a function of $\vect x'  = \sum_{n=1}^{2N} x'_n \vect e_n$, where the real and imaginary parts of $\vect x$ are stacked in, the constraint is reformulated as 
\begin{align}
 x'_n \leq 1/\sqrt{2} \text{ and } -  x'_n \leq 1/\sqrt{2}, \:\: n=1,2,...2N.
\label{eq:relaxed_constraint}
\end{align}

\subsection{Product-Maximization (PM)}
Maximizing the sum may lead to maximizing the expression for the user with the highest value at the cost of other users.

Maximizing the product seems to be a fairer method, since the product can be maximized only if all the values of all the users contribute considerably.
Thus, the relaxed optimization problem reads as
\begin{align}
&\max_{\vect x'} \det(\matt P) \nonumber \\
&\text{  s.t. }   x'_n \leq 1/\sqrt{2}
 \text{ and } - x'_n \leq 1/\sqrt{2}, n=1,2,...2N. 
\end{align}

For this optimization problem, we resort to the gradient projection algorithm \cite{NLP} to fulfill the constraint in (\ref{eq:relaxed_constraint}). To this end, we need to find the derivative expression of the cost function with respect to $\vect x'$.
The gradient is given by
%\begin{align}
%\frac{\partial \det(\matt P)}{\partial  x'_n} &= \det(\matt P) \tr \left(\matt P^{-1} \frac{\partial \matt P}{\partial  x'_n}\right) \nonumber \\
%&=  2 \det(\matt P) \tr \left(\matt P^{-1}\left(\matt A \frac{\partial \matt A}{\partial  x'_n} -  \matt B \frac{\partial \matt B}{\partial  x'_n}\right)\right) \nonumber \\
%&=  2 \det(\matt P) \tr \Bigg(\matt P^{-1}\Big(\matt A  \diag \left( \matt C \vect e_n  \vect{s_{\Re}}^{\T} + \matt D \vect e_n  \vect{s_{\Im}}^{\T} \right) \nonumber \\
%&  -\matt B  \diag \left( \matt C \vect e_n  \vect{s_{\Im}}^{\T} - \matt D \vect e_n  \vect{s_{\Re}}^{\T} \right)\Big) \Bigg) \nonumber \\
%&= 2 \sum_{m=1}^{M} \frac{\det \left(\matt P \right)}{\vect e_m^{\T} \matt P \vect e_m} \cdot \nonumber \\
%&\Big( \vect e_m^{\T} \matt A \vect e_m  \vect e_m^{\T}\left( \matt C \vect e_n  \vect{s_{\Re}}^{\T} + \matt D \vect e_n  \vect{s_{\Im}}^{\T} \right) \vect e_m \nonumber \\
%&- \vect e_m^{\T} \matt B \vect e_m  \vect e_m^{\T}\left( \matt C \vect e_n  \vect{s_{\Im}}^{\T} - \matt D \vect e_n  \vect{s_{\Re}}^{\T} \right) \vect e_m\Big) \nonumber \\
%&= 2 \Big(\vect s_{\Re}^{\T} \adj\left( \matt P \right) \left( \matt A \matt C + \matt B \matt D \right) \nonumber \\
%&+ \vect s_{\Im}^{\T} \adj\left( \matt P \right) \left( \matt A \matt D - \matt B \matt C \right)\Big) \vect e_n.
%\end{align}
%Thus, we get
\begin{align}
\frac{\partial \det(\matt P)}{\partial \vect x'} &= 2 \matt C^{\T} \adj \left( \matt P \right) \left(\matt A \vect s_{\Re} - \matt B \vect s_{\Im} \right) \nonumber \\
&+ 2 \matt D^{\T} \adj \left( \matt P \right) \left(\matt A \vect s_{\Im} + \matt B \vect s_{\Re} \right).
\end{align}

\subsection{Gradient Projection Algorithm}
The used algorithm is summarized in Algorithm \ref{table:gpa}. The initial value of $\vect x'_{(0)}$ depends on the choice of $\matt W$, where $\matt W$ is chosen as zero-forcing (ZF) precoder $$\matt W= \matt H^{\He}\left(\matt H \matt H^{\He} \right)^{-1}. $$
\begin{algorithm}
\caption{Gradient Projection Algorithm}
\label{table:gpa}
\begin{algorithmic}[1]
\State{Iteration step  $\mu=\mu_0$, Tolerable error $\epsilon = 10^{-6}$}
\State{\textbf{Initialization}$ \: \vect x'^{\T}_{(0)} = \begin{bmatrix}
\Re \{\matt W \vect s\}^{\T} & \Im \{\matt  W \vect s\}^{\T}
\end{bmatrix}^{\T},$ $i = 0$}    \\   
\textbf{if} {$\begin{cases}
 x'_{(0), n} > 1/\sqrt{2} & \text{\textbf{then} } x'_{(0), n}= 1/\sqrt{2}\\
 - x'_{(0), n} > 1/\sqrt{2} & \text{\textbf{then} }  x'_{(0), n} = -1/\sqrt{2}
 \end{cases}$}
\Repeat \\{ $\vect x'_{\left ( i+1 \right )}=\vect x'_{\left ( i\right )}+\mu\left ( \frac{\partial \det( \matt P_{\left ( i\right )})} {\partial \vect x'}\right )$\\
\textbf{if} {$\begin{cases} x'_{\left ( i+1 \right ),n} > 1/\sqrt{2} & \text{\textbf{then} } x'_{\left ( i+1 \right ),n} = 1/\sqrt{2} \\ -x'_{\left ( i+1 \right ),n} > 1/\sqrt{2} &  \text{\textbf{then} } x'_{\left ( i+1 \right ),n} =- 1/\sqrt{2} \end{cases}$}\\
\textbf{if} {$\begin{cases} \det(\matt P_{(i+1)}) < \det(\matt P_{(i)} \\
\matt A_{m,m} < 0 \\
\matt P_{m,m} <0, m=1,...,M  \end{cases} \text{\textbf{then} } \mu=\mu/2$}
\\$i=i+1$}
\Until{$\left(\det(\matt P_{(i+1)}) - \det(\matt P_{(i)})\right)/\det(\matt P_{(i)}) \leq \epsilon $}
\end{algorithmic}
\end{algorithm}

The iteration step start value is denoted by $\mu_0$. If the iteration step is very large such that the cost function decreases instead of increasing or the elements of $\matt P$ or $\matt A$ become negative, the step size has to be reduced in order to ensure the algorithm convergence. This iteration step optimization is performed in step 7.

\section{Existing linear precoders}
\label{sec:linear_precoders}
\subsection{WF precoder}
This precoder design was introduced in \cite{Joham2005}. It is based on the MMSE criterion and is given by
\begin{align*}
\matt W_{\text{WF}} &=\frac{1}{f_{\text{WF}}}\left(\matt H^{H}\matt H+\frac{M \matt I_{N}}{E_{\textrm{tx}}}\right)^{-1}\matt H^{H}, \nonumber \\
f_{\text{WF}} &=\sqrt{\frac{\sigma_s^2}{E_{\textrm{tx}}}\textrm{tr}\left ( \left ( \matt H^{H}\matt H + \frac{M \textbf{I}_{N}}{E_{\textrm{tx}}})\right )^{-2}\matt H^{H}\matt H \right )}.
\end{align*}
The transmit vector reads then as $\vect x_Q = \mathcal{Q}\left(\matt W_{\text{WF}} \vect s  \right)$. To fulfill the power constraint the transmit vector $\vect x_Q$ is scaled by the factor $\sqrt{\frac{E_{\text{tx}}}{N}}$, which ensures equal power allocation at the antennas.
\subsection{WFQ precoder}
This precoder design was presented in \cite{Mezghani2009}. It is an MMSE precoder that takes into account the quantization effects based on the linear covariance approximation. The precoder is expressed by
\begin{align*}
&\matt W_{\text{WFQ}}\!\!=\!\!\frac{1}{f_{\text{WFQ}}}\!\!\left (\matt H^{H}\matt H\!-\!\rho_{q}\textrm{nondiag} \left (\matt H^{H}\matt H  \right ) \!\!+\frac{M\matt I_{N}} {E_{\textrm{tx}}}\right )^{-1}\!\!\!\matt H^{H}, \nonumber \\
&f_{\text{WFQ}}=\sqrt{\frac{\sigma_s^2 (1-\rho_q)}{E_{\textrm{tx}}}} \cdot \nonumber \\
&\sqrt{\textrm{tr}\left (\left (\matt H^{H}\matt H\!-\!\rho_{q}\textrm{nondiag} \left (\matt H^{H}\matt H  \right ) \!+\!\frac{M \matt I_{N}} {E_{\textrm{tx}}}\right )^{-2} \!\! \matt H^{H}\matt H \right) },
\end{align*}
where $\rho_q = 1-\frac{2}{\pi}$. This  precoder design consists of two stages: the digital precoder $\matt W_{\text{WFQ}}$ and the analog precoder $\matt D_{\text{WFQ}}=\sqrt{2/\pi}~ \diag(\matt W_{\text{WFQ}}\matt W_{\text{WFQ}}^{\He})^{1/2} $. The analog precoder is a diagonal real-valued matrix to assign each antenna a desired amount of power and to optimize the quantization levels. So we end up with $\vect x_Q = \mathcal{Q}\left(\matt W_{\text{WFQ}} \vect s\right)$ that has to be multiplied with $\matt D_{\text{WFQ}}$ before transmitting. This leads to unequal power allocation at the BS antennas.

\section{Simulation Results}
\label{sec:simresults}
All the simulation results are averaged over 500 channel realizations. The used modulation scheme is QPSK, where $\sigma^2_s =1$,  with $N_b=10^4$ transmit symbols per channel use. We compare our proposed design PM with existing linear precoding techniques WF and WFQ in terms of the uncoded BER and the mutual information (MI) for $N=32$ and $M=4$. The MI is calculated numerically based on the toolbox proposed in \cite{UM2012}. The ideal case, where the WF precoder is used and no quantization is performed, is denoted by "WF, unq.".

\begin{figure}[!t]
\centering
\psfrag{BER}[][]{Uncoded BER}
\psfrag{SNR (dB)}[][]{$E_{\text{tx}}$(dB)}
\psfrag{QWF}[][]{\:\:\:\: WFQ}
\psfrag{MMSE}[][]{ WF}
\psfrag{MBER}[][]{ PM}
\psfrag{Ideal case}[][]{ \: WF, unq.}
\includegraphics[width=\columnwidth]{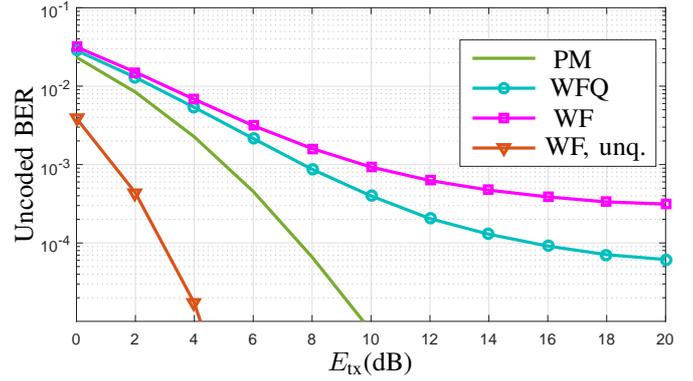}

\caption{BER performance for a MU-MISO system with $N=32$ and $M=4$.}
\label{fig:ber_32_4}
\end{figure}

From  Fig. \ref{fig:ber_32_4} we can see that
the proposed mapping method outperforms the existing linear precoders in terms of uncoded BER. At uncoded BER of $10^{-3}$ we achieve a gain compared to WFQ of 3dB. In the WFQ design, unequal power allocation at the antennas is performed. This requires a number of power amplifiers (PAs) equal to the number of antennas to adjust the power for each antenna.
In our proposed method the power at each antenna is equal which allows to run the PA in the saturation region and thus efficiently use the energy.

\begin{figure}[!t]
\centering
\psfrag{MI}[][]{MI}
\psfrag{SNR (dB)}[][]{$E_{\text{tx}}$(dB)}
\psfrag{QWF}[][]{\:\:\:\: WFQ}
\psfrag{MMSE}[][]{ WF}
\psfrag{MBER}[][]{  PM}
\psfrag{Ideal case}[][]{ \: WF, unq.}
\includegraphics[width=\columnwidth]{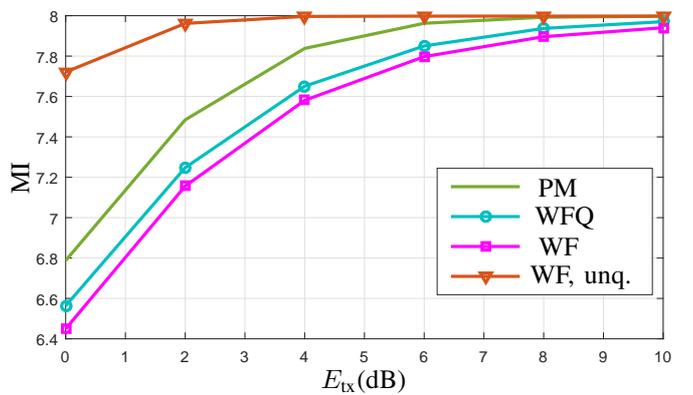}
\caption{MI performance for a MU-MISO system with $N=32$ and $M=4$.}
\label{fig:mi_32_4}
\end{figure}

In Fig. \ref{fig:mi_32_4} the MI for the different precoder designs are plotted as function of the transmit power. The gain in MI is less significant compared to the uncoded BER. This means that the proposed method requires less perfomant codes to achieve the capacity.

\begin{table}[!t]
\renewcommand{\arraystretch}{1.3}
% if using array.sty, it might be a good idea to tweak the value of
 %\extrarowheight as needed to properly center the text within the cells
\caption{Complexity and performance of the PM method}
\label{table_example}
\centering
% Some packages, such as MDW tools, offer better commands for making tables
% than the plain LaTeX2e tabular which is used here.
\begin{tabular}{|c|c|c|c|}
%\multicolumn{4}{c}{$N=32$, $M=4$} \\
\hline
$\epsilon$ & average nb. of iterations & SNR @ BER of $10^{-3}$ & MI \\
\hline
$10^{-3}$ & 11 & 5,3dB & 7,92 bpcu\\
\hline
$10^{-4}$ & 18 & 5,45dB & 7,92 bpcu\\
\hline
$10^{-6}$ & 43 & 5,08dB & 7,92 bpcu\\
\hline
\end{tabular}
\label{table2}
\end{table}

Additionally, the complexity of the PM method is studied in terms of the average number of iterations needed to get one optimal solution for $\vect x$ in Table \ref{table2}. As can be drawn from the table, the required number of iterations decreases with larger tolerable error $\epsilon$. We can go to around 10 iterations per algorithm run without degrading much the uncoded BER and the MI.

\section{Conclusion}
\label{sec:conc}
We presented a novel precoding technique based on the MBER criterion. Instead of designing a precoder we design the transmit output vector that fulfills the relaxed constraint of QPSK set to minimize the distortions due to the 1-bit quantization at the transmitter based on the MBER criterion. This method gives promising results compared to the existing linear precoding techniques. Although equal power allocation at the BS antennas is performed, we achieve a significant gain of 3dB at BER of $10^{-3}$ compared to precoders that allow unequal power allocation at the antennas. Furthermore, the PA can be run in the saturation region to get more energy efficient systems. However, these advantages are achieved with higher complexity of running a nonlinear optimization problem for each input. However, a LUT based implementation is possible for systems with small number of users.

\newpage

\bibliographystyle{IEEEtran}
\bibliography{IEEEabrv,refs}

\end{document}